\documentstyle[12pt]{article}

\begin{document}

\title{Determination of critical current density in melt-processed HTS bulks
from levitation force measurements}

\author{A. A. Kordyuk\thanks{E-mail: kord@imp.kiev.ua}, V. V. Nemoshkalenko, R. V. Viznichenko\\
Institute of Metal Physics, Kyiv, Ukraine\\
\\
T. Habisreuther, W. Gawalek\\
Institut f\"ur Physikalische Hochtechnologie, Jena, Germany}

\date{March 31, 1999}

\maketitle

\begin{abstract}
A simple approach to describe the levitation force measurements on
melt-processed HTS bulks was developed. A couple of methods to determine
the critical current density $J_c$ were introduced. The averaged $ab$-plane
$J_c$ values for the field parallel to this plane were determined. The first
and second levitation force hysteresis loops calculated with these $J_c$
values coincide remarkably well with the experimental data.
\end{abstract}

\section{Introduction}
Superconducting systems with magnetic levitation have long been known and the 
discovery of high temperature superconductors (HTS) highly stimulated their
investigation but a real interest in them for large scale applications appears
only with the successful development of the melt-processed (MP) technology
\cite{1}. The use of MP HTS in large scale systems such as flywheels for
energy storage, electric motors and generators, permanent magnets, etc. is
the most promising HTS application now \cite{2}. In this applied region the
levitation force measurements can be considered in two roles: as an information
source to know more about levitation systems and as a quick technique to test
HTS samples \cite{3}. In many earlier works \cite{4,5} it has been shown that
the forces between a PM and HTS sample are closely related with HTS
magnetization curves. Vertical levitation force versus vertical distance
$F_z(z)$ is the nearest analog to $M(H)$ dependencies with their major and
minor hysteresis loops but the complexity of a field configuration in such
large scale PM-HTS systems makes it very difficult to directly correlate them
in general case. The problem can be solved by numerical approaches and some 
of them have been successfully used \cite{6}. The numerical approaches are
undoubtedly useful to evaluate the real system parameters but usually need too
much computer resources to be applicable to direct HTS sample investigation.
To perform such an investigation an analytical evaluation is more wished for.

Two limiting cases of HTS structure have been considered as analytical to
calculate the dynamic parameters of an idealized system, a point magnetic
dipole over an infinite flat superconductor. The first one is the case of
`granular superconductor' which can be modeled by a set of small isolated
superconducting grains \cite{7}. It was shown the usual granular HTS obtained
by standard sintering techniques can be described very well within this model
\cite{7,8}. The second one is the case of an `ideally hard superconductor'
\cite{9,10,11,12}. It was shown recently \cite{9,10,11} that dynamics in a
wide variety of levitation systems can be described in terms of surface
screening currents which screen alternating magnetic field component due to
PM displacements like it would be for an ideally hard superconductor, a
superconductor with infinite pinning forces. For an infinite flat
superconductor the frozen-image method was introduced \cite{9} as an
illustration of simple analytical calculation of forces acting in the system.
A perfect agreements with experiments were found by us for small PM resonance
oscillations frequencies \cite{10} and, recently, by Hull and Cansiz
\cite{13}, for both vertical and lateral force components.

\section{Approach description}
The feasibility of the `ideally hard superconductor' approach is that the
penetration depth $\delta$ of alternating magnetic field is much less than
system dimensions \cite{9,10}. To calculate the stiffness or resonance
frequencies the limit $\delta \to 0$ can be used, but it was shown that taking
into account the finite values of $\delta$ it is possible to calculate the
a.c. loss \cite{11} and even recover critical current density profiles within
$\delta$ depth from a.c. loss measurements \cite{12}. In this paper we present
such an approach of levitation force calculation (including its hysteresis
behavior) for superconductor with finite values of critical current density
$J_c$ and simple methods to obtain $J_c$ values from levitation force
experimental data.  

A PM placed over ideally hard superconductor induces at its surface the
screening currents ${{\bf j} = (c/4\pi) {\bf n} \times {\bf b}_r}$, where
${\bf n}$ is the surface normal and ${\bf b}_r$ is the tangential magnetic
field component at the surface (the normal component $b_n$ at the surface
is zero). From the symmetry, for an infinite flat surface \cite{10} 

\begin{eqnarray}
{{\bf b}_r = 2 {\bf b}_{ar}},	\label{1}
\end{eqnarray}

\noindent
where ${\bf b}_a$ is the variation of the PM magnetic field ${\bf B}_a$ due
to its displacements in respect to initial field cooled (FC) position:
${\bf b}_a = {\bf B}_a - {\bf B}_{aFC}$ \cite{10}. For the $z$-axial
symmetric configuration ${\bf r} = (r sin\theta, r cos\theta, z)$ where
only $j_\theta$ component is induced, for the vertical force acting
on PM from the screening currents one can write

\begin{eqnarray}
F_{id} = \int\limits^{\infty}_{0} r b^2_{ar}(r) dr. \label{2}
\end{eqnarray}

This is ideal force which can be readily calculated just from known tangential
component of PM field. The Eq.(\ref{2}) is obtained from zero-depth screening
currents approach that we will call a zero approximation of real PM-HTS systems.
Within this approximation any configuration of such systems can be calculated
numerically \cite{14} but to describe hysteresis phenomena next-order 
approximations have to be considered.

In the second stage (a first approximation) we will examine a model where:
(i) $\delta$ is finite but still much less than system dimensions $L$, (ii)
the critical state model is applicable to these samples, and (iii) critical
current density is constant. The applicability of the critical state model 
to melt-processed HTS has been proven in many experiments \cite{3,11,12} and
is quite acceptable here. The first condition on $\delta$ can be written as

\begin{eqnarray}
\delta(r) \ll B_{ar}(r)\left({\frac{dB_{ar}(r)}{dr}}\right)^{-1} \sim L,
\label{3}
\end{eqnarray}

\noindent
and, because of $\delta \propto b_r$, can be satisfied anyway limiting the
minimum distance between PM and HTS surface. One can estimate $L \approx z
+ d/2$, where $z$ is, here and below, the distance between PM and HTS surface
and $d$ is the PM thickness.

As for condition on $J_c$, it is far from reality by itself since the critical
current density usually depends on both magnetic field and space coordinate.
But as it will be shown below we can accept this condition for levitation
force measurements. This just means that in the next relation for $j$, the
surface density of screening currents,

\begin{eqnarray}
j(r, z) = \delta(r, z) J_c = \frac{c}{2\pi} b_{ar}(r, z), \label{4}	
\end{eqnarray}

\noindent
$J_c$ can be treated as a coefficient between $j$ and $\delta$, a coarse-grained
flux penetration depth averaged over $L$ scale. The $j(r)$ function does not
depend on field history but only on PM position $z$ in the same way as
$b_{ar}(r)$, the distribution at the HTS surface of the PM field variation,
that after cooling is a function of $r$ and $z$. Thus, in the considering
approximation, the function $\delta(r, z)$ formally does not depend on field
history but means the flux penetration depth at the first PM descent only.

Next, if we use a protocol of PM motion according to which it moves between
two points: the initial or FC point $z_{max}$ that is included in
$b_{ar}(r, z)$ function as a condition $b_{ar}(r, z_{max}) = 0$ 
($z_{max} = \infty$ for ZFC case), and the lowest point $z_{min}$, the current
distributions in the depth $z$ of superconductor are the following. After the
PM first stop and beginning to go up (the first ascent), the depth of the layer
where currents flow remains constant and equal to its maximum value
$\delta_{max} \equiv \delta(r, z_{min})$ but there are two regions with
opposite currents. The opposite flowing current penetrates from the top at the
depth $\delta_{\uparrow}$ that can be obtained from (4)

\begin{eqnarray}
\delta_{\uparrow}(r, z, z_{min}) = \frac{1}{2} (\delta(r, z_{min}) -
\delta(r, z)). \label{5}
\end{eqnarray}

Its maximum value is $\delta_{\uparrow max} \equiv \delta_{\uparrow}(r,
z_{max}, z_{min}) = \delta(r, z_{min})/2$, so during the second descent 
there are three regions with $+J_c$ for $0 < \zeta < \delta_{\downarrow}$ ,
$-J_c$ for $\delta_{\downarrow} < \zeta < \delta_{\uparrow max}$ , and $+J_c$
for $\delta_{\uparrow max} < \zeta < \delta_{max}$ , where $\delta_{\downarrow}
(r, z) = \delta(r, z)/2$ also does not depend on $z_{min}$. If one can neglect 
flux creep for times greater than the descent-ascent time, any other ascents
are equal to the first one and any other descents are equal to the second one.
Any other current distributions for other protocols, for example to describe
minor hysteresis loops, can also readily be obtained within the scheme above.

Applying this scheme to calculate the vertical forces during the first descent
$F(z)$, the first and the next ascents $F_{\uparrow}(z, z_{min})$ and the
second and the next descents $F_{\downarrow}(z, z_{min})$ one can write

\begin{eqnarray}
F(z) = \frac{2\pi}{c} J_c \int\limits^{\infty}_{0} r dr
\int\limits^{\delta(r,z)}_{0}d\zeta b_{ar}(r, z+\zeta). \label{6}	
\end{eqnarray}

\begin{eqnarray}
F_{\uparrow}(z, z_{min}) = \frac{2\pi}{c} J_c \int\limits^{\infty}_{0} r dr
\left[\int\limits^{\delta_{max}}_{\delta_{\uparrow}}d\zeta
- \int\limits^{\delta_{\uparrow}}_{0}d\zeta \right]
b_{ar}(r, z+\zeta). \label{7}
\end{eqnarray}

\begin{eqnarray}
F_{\downarrow}(z, z_{min}) = \frac{2\pi}{c} J_c \int\limits^{\infty}_{0} r dr
\left[\int\limits^{\delta_{max}}_{\delta_{max}/2}d\zeta
- \int\limits^{\delta_{max}/2}_{\delta/2}d\zeta
+ \int\limits^{\delta/2}_{0}d\zeta \right]
b_{ar}(r, z+\zeta). \label{8}
\end{eqnarray}

The functions $\delta(r,z)$ depend on $J_c$ in according to the above
equations ((\ref{4}), (\ref{5}) and below) and for $J_c \rightarrow \infty$
all these forces become equal to $F_{id}(z)$.

Remaining within the condition (\ref{3}) we can approximate the integrals
over $z$ from the formulas (\ref{6})-(\ref{8}) by multiplying the depth of
the layer where current flows by the field bar in its center. It is easy to
show that within the above approximation the formula (\ref{6}), for example, 
can be rewritten as

\begin{eqnarray}
F(z) = \int\limits^{\infty}_{0} r b_{ar}(r,z) b_{ar}\left(r,z +
\frac{\delta}{2} \right) dr,
\label{9}	
\end{eqnarray}

\noindent
which highly increases the calculation speed.

\section{Experiment}
To check the applicability of the above consideration to real MP HTS we used
a standard experimental setup on levitation force measurements \cite{3}.
The SmCo$_5$ disk shape PM was 15 mm in diameter and 8 mm in thickness
(the effective thickness with ferromagnetic holder that was evaluated from
real PM field configuration was 12.7 mm) with averaged axial magnetization of 
$4\pi M$ = 9236 G (the field measured by Hall probe in its center at the
distance of 0.8 mm from its bottom surface was 3350 G). The magnetic field
of the PM was calculated as field of a coil with the same dimensions and with
lateral surface current density $J = c M$. All measured samples were
melt-processed HTS of 30$\pm$0.5 mm in diameter and 17.5$\pm$0.5 mm in
thickness. The distance $z$ between PM bottom surface and HTS top surface
varied from $z_{max}$ = 400 mm (that can be considered as ZFC case) to its
minimum value $z_{min}$ = 0.5 mm. The minimum step of PM motion was 75 mm.
The accuracy of force detecting was 15 mN. Within this accuracy the
experimental data were reproducible for every sample. Fig.1 represents the
first and second hysteresis loops (the first and second descent and ascent)
for two samples.

Within the above approximation we have only one parameter, $J_c$, for forces
(\ref{6})-(\ref{8}) (or (\ref{9}) and analogous ones) to be fitted to the
experimental ones $F_{exp}(z)$. To do this, we have to choose one of these
functions and one point $z_i$, and solve the equation 

\begin{eqnarray}
F(J_c, z_i) = F_{exp}(z_i).	\label{10}
\end{eqnarray}

The forces calculated from formulas (\ref{6})-(\ref{8}) with the $J_c$ values
obtained from (\ref{10}) in $z_{min}$ point are also represented in Fig.1
by solid lines. The forces calculated from the formula (\ref{9}) and from
analogous ones practically coincide with the above in the $F(z)$ plot scale.
A good agreement between the experimental and calculated $F(z)$ dependencies
demonstrates the above approximation is correct.

Nevertheless, the discrepancy between the experimental and calculated forces
still exists and is lager than the experimental accuracy. One of the most
likely reasons is a variation of $J_c$ with depth and field. Fig.2 shows the
values of $J_c$ versus maximum value of $B_r(z)$ at the HTS surface for two
HTS samples. The data were obtained by solving Eq.(\ref{10}). Open symbols 
represent the solution for the function (\ref{6}), and solid symbols represent
the solution for the function (\ref{9}). The solid line in Fig.2 with respect
to right axis represents the dependence of $B_{rmax}(z)$. For a perfectly
uniform sample with $c$-axis exactly perpendicular to the surface such 
a dependence of $J_c(B_{rmax})$ would be uniquely determined by the dependence
of $J^{ab}_{c}(B_{ab})$, the critical current density flowing in $ab$-plane
versus the magnetic field parallel to this plane. But for real melt processed
samples, it is more reasonable to assume that the dependencies of 
$J_c(B_{rmax})$ in Fig.2 are mostly caused by space variations of critical
current density. The steep slope of the curves in Fig.2 at low field, which
caused the maximum in the apper curve, is rellated with the finite diameter
of the HTS samples and shows the lower field limit for given configuration.

\section{Another simple method}
There is a possibility here to introduce a visual simple method to evaluate
$J_c$. It is understandable that in spirit of the above consideration a shift
$\Delta z$ (see Fig.1) of the first descent experimental curve with respect
to ideal one has to be proportional to an average penetration depth. From the
condition $F_{id}(z + \Delta z) = F(J_c, z)$ and Eq.(\ref{9}) one can readily
obtain

\begin{eqnarray}
\delta \approx 4\Delta z, \textrm{ or }
J_c \approx \frac{c}{8\pi} \frac{b_{ar}}{\Delta z}.
\label{11}
\end{eqnarray}

The values of $J_c$ evaluated in such way are also represented in Fig.2 and
show a good agreement with ones determined before. The experimental error
$\sigma_{J_c}$ that is shown here was 
estimated from the formula $\sigma_{J_c}/J_c = \sigma_F (dF/dz)^{-1} /
\Delta z$ which assumes the maximum error is caused by the force measuring:
$\sigma_F \approx$ 30 mN. 

\section{Conclusions}
In summary, we have considered the approach, which we call the "first
approximation", to describe levitation force data. The term "first" implies
that we consider a case in which such parameters as flux penetration depth
$\delta$ or normal component of magnetic field at HTS surface $b_n$ are already
not zero, as it is for ideally hard superconductor \cite{9}, but small enough:
$\delta \ll L$, $b_n \ll b_r$. Within this condition the methods to calculate
$J_c$, the critical current density, which we have introduced in the paper are
exact. Remarkably, the approach works well even beyond this condition, when
$\delta \sim L$, $b_n \sim b_r$. In this region the methods become empirical.
The $J_c$ value that can be obtained by the methods is averaged over $L$ scale
critical current density in $ab$-plane for field parallel to this plane:
$J_c = \langle J_{c}^{ab}({\bf B} \parallel ab) \rangle$. $L$ scale depends
on the size of a magnet we use.

\section{Acknowledgments}
The authors would like to thank J.~R.~Hull for helpful discussions and
T.~Strasser for assistance with the experimental setup.

\newpage
\section{Figures Captions}
Fig. 1. Experimental (symbols) and calculated (solid lines) data on the first
and second hysteresis loops of the vertical levitation force vs. distance $z$
between PM and HTS surface. Dashed line represents the force for an ideal
superconductor.
 \\ [0.5 cm]
Fig. 2. The values of averaged critical current density versus maximum value
of $B_r$, the magnetic field tangential component at the HTS surface, obtained
by different methods. The solid line with respect to right axis represents the
dependence of $B_{rmax}(z)$.


\begin{thebibliography}{99}
\bibitem{1}
D. A. Cardwell, Mat. Sci. Eng. B, 53, 1 (1998).
\bibitem{2}
J. R. Hull, IEEE Spectrum 34,  20 (1997).
\bibitem{3}
G. Fuchs, P. Stoye, T. Staiger, G. Krabbes, P. Schatzle, W. Gawalek,
P. Gornert, and A. Gladun, IEEE Trans. Appl. Supercond. 7, 1949 (1997).
\bibitem{4}
E. H. Brandt, Appl. Phys. Lett. 53, 1554 (1988).
\bibitem{5}
F. C. Moon, K.-C. Weng, and P.-Z. Chang, J. Appl. Phys. 66, 5643 (1989).
\bibitem{6}
T. Sugiura, H. Hashizume, and K. Miya, Int. J. Appl. Electromag. Mater. 2,
183 (1991).
\bibitem{7}
A. A. Kordyuk and V. V. Nemoshkalenko, Appl. Phys. Lett. 68, 126 (1996).
\bibitem{8}
A. A. Kordyuk and V. V. Nemoshkalenko, J. Supercond. 9, 77 (1996).
\bibitem{9}
A. A. Kordyuk, J. Appl. Phys. 83, 610 (1998).
\bibitem{10}
A. A. Kordyuk, Metal. Phys. Adv. Tech. 18, 249 (1999).
\bibitem{11}
A. A. Kordyuk, V. V. Nemoshkalenko, R. V. Viznichenko, and W. Gawalek,
Mat. Sci. Eng. B 53, 174 (1998).
\bibitem{12}
A. A. Kordyuk, V. V. Nemoshkalenko, R. V. Viznichenko, and W. Gawalek,
Physica C 310, 173 (1998).
\bibitem{13}
J. R. Hull and A. Cansiz, J. Appl. Phys. (1999) to be published.
\bibitem{14}
V. V. Vysotskii and V. M. Pan, Inst. Phys. Conf. Ser. 158, 1667 (1997).
\end{thebibliography}
\end{document}